\documentclass[aps,pra,preprint,groupedaddress,showpacs]{revtex4-1}
\usepackage{graphicx}
\usepackage{bm}
\begin{document}

\title{Spin Nutation Induced by Atomic Motion in a Magnetic Lattice}
\author{Y. Kobayashi}
\author{Y. Shiraishi}
\author{A. Hatakeyama}
\email[]{hatakeya@cc.tuat.ac.jp}

\affiliation{Department of Applied Physics, Tokyo University of Agriculture and Technology, Koganei, Tokyo 184-8588, Japan}

\date{\today}

\begin{abstract}
An atom moving in a spatially periodic field experiences a temporary periodic perturbation and undergoes a resonance transition between atomic internal states when the transition frequency is equal to the atomic velocity divided by the field period. We demonstrated that spin nutation was induced by this resonant transition in a polarized rubidium (Rb) atomic beam passing through a magnetic lattice. The lattice was produced by current flowing through an array of parallel wires crossing the beam. This array structure, reminiscent of a multiwire chamber for particle detection, allowed the Rb beam to pass through the lattice at a variety of incident angles. The dephasing of spin nutation was reduced by varying the incident angle.
\end{abstract}
\maketitle

\section{Introduction}
  
Coherent interactions between atoms or quantum systems and resonant electromagnetic radiation are of great importance in modern physics. They are widely used for preparing and detecting quantum states, providing indispensable techniques in many research fields, including rapidly progressing quantum information studies such as quantum entanglement~\cite{Rai01}. The Rabi oscillation is a fundamental coherent phenomenon in which resonant monochromatic radiation periodically exchanges the populations between atomic internal states. The Rabi oscillation is observed when an ensemble of atoms is subjected to a sinusoidally varying perturbation from the radiation. This perturbation, however, does not necessitate electromagnetic radiation. We investigated coherent spin nutation by resonant excitation in an atomic beam without electromagnetic radiation using a periodic magnetostatic field (``magnetic lattice''), through which the atomic beam passed.
The magnetic resonance occurred inside the lattice when the frequency of the periodic perturbation experienced by the atoms was equal to the transition frequency. 

This resonance transition has been extensively studied using fast ion beams in periodic crystal fields and has often been referred to as ``resonant coherent excitation'' (RCE)~\cite{Dat78}. Recent progress achieved using high-energy beams of highly charged ions showed the potential for applications in high-resolution spectroscopy or even manipulation of the internal states of these ions in the X-ray region of keV or $10^{18}$~Hz~\cite{Azu99, Kon06, Nak08, Nak09}, in which the order was determined by the ion velocity ($\sim10^8$~m/s) and the lattice constant ($\sim10^{-10}$ m). The principle of RCE is quite general and has recently been extended to low-energy experiments~\cite{Hat05, Hat08, Kob09}, with the objective of developing a new type of atomic manipulation technique, particularly near the surface. These experiments demonstrated that magnetic resonance was induced by atomic motion through an artificial magnetic lattice, referred to as ``motion-induced resonance'' (MIR). Sharp resonance lines were obtained, and their widths were primarily determined by the transit time through the lattice ~\cite{Kob09}. However, in both high-energy and low-energy experiments, coherent Rabi-type oscillation of the population has not been observed directly~\cite{Nak08}. Our observation of spin nutation is the first direct demonstration of coherent population transfer induced by atomic motion through a periodic field.

In our experiment, an optically spin-polarized rubidium (Rb) atomic beam entered the magnetic lattice with a velocity of a few hundred meters per second. The spatial period of the magnetic field was 2~mm, and the frequency that the atoms experienced was on the order of a hundred kHz, corresponding to a transition between ground state Zeeman sublevels that have been split by an external longitudinal magnetic field. The resultant population transfer was measured after the atoms exited the lattice.

The magnetic lattice was produced by current flowing through an array of parallel wires crossing the beam. The wire array was formed by approximately seven layers of planar arrays of parallel wires. This structure, reminiscent of a multiwire-chamber-type particle detector \cite{Cha68}, allowed a variety of incident angles. This was unattainable with our previous version of the magnetic lattice, which was formed using a stack of printed circuit boards \cite{Kob09}, and each atom was only able to travel through one of the gaps between two adjacent boards. This degree of freedom of the incident angle can be used to reduce the dephasing of spin nutation caused by different field amplitudes for different incident positions. The usefulness of this method was first recognized in a previous RCE experiment ~\cite{Kon06}, and in our experiment, the dephasing of nutation signals actually decreased when the incident angle of the beam was varied.

In this paper, the experimental setup is described in Sec.~II, followed by a detailed description of the magnetic lattice in Sec.~III. Results from magnetic resonance spectra and spin nutation are presented and discussed in Sec.~IV. Finally, conclusions are drawn in Sec.~V.

\section{Experiment}

\begin{figure}
\includegraphics{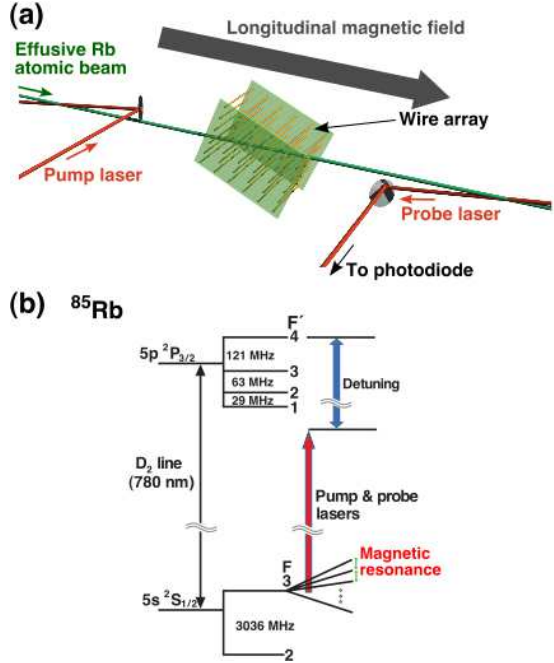}
\caption{(color online). (a) Schematic of the experimental setup. (b) Relevant energy levels of $^{85}$Rb. The Zeeman sublevels of the ground state $F=3$ level, split by a longitudinal magnetic field for magnetic resonance, are shown. The laser detuning is also shown.}
\end{figure}

A schematic of the experimental setup is shown in Fig.~1(a).
Rb atoms emerged from an oven held at $\sim200^{\circ}$C and were collimated by two holes that were 2.0~mm in diameter separated by $100$~mm. The Rb atoms were optically polarized before entering the magnetic lattice. The energy level diagram of $^{85}$Rb, relevant to optical pumping and probing, is shown in Fig.~1(b). A circularly polarized pump laser (wavelength: 780.24~nm), nearly collinear to the atomic beam, polarized the $^{85}$Rb atoms in the $F=3$ ground state by optical pumping through the $F=3 \rightarrow F'=4$ cyclic transition. A longitudinal magnetic field up to 0.2~mT was applied using a pair of coils along the atomic beam direction ($z$ direction). The polarized beam interacted with the magnetic lattice after collimation using a slit with a 2-mm gap. The lattice had a spatial period of $a=2.0$~mm (details explained in the next section). It rotated for varying the incident angle of the beam.
Magnetic resonance transitions occurred between the Zeeman sublevels of the $F=3$ state that was split by the longitudinal magnetic field, resulting in spin nutation as the atoms passed through the lattice. After the atoms exited the lattice, the $z$ component of the atomic spin $\langle F_z\rangle$ was detected using a probe laser that had the same frequency as the pump laser and was also nearly collinear to the atomic beam through the $F=3 \rightarrow F'=4$ transition. The polarization of the probe laser was modulated between left and right circular polarizations at 42~kHz using a photoelastic modulator, and the resultant laser absorption signals were demodulated using a lock-in amplifier. The magnetic resonance spectra were recorded as a function of the strength of the longitudinal magnetic field, with the strength of the lattice field fixed. Thus, they were a function of the Zeeman splitting. Spin nutation signals were measured as a function of the lattice field, with the longitudinal magnetic field fixed at resonance. The entire vacuum chamber and the coils producing the longitudinal magnetic field were enclosed in a magnetic shield.

The laser beams also worked as a velocity selector in the pumping and probing processes. They selectively detected atoms within a narrow velocity range from a broad Maxwell--Boltzmann distribution of the effusive atomic beam using the Doppler effect. The laser frequency was stabilized to one of the $^{85}$Rb resonance lines of the saturation absorption spectrum and was further shifted using an acousto-optic modulator. The resultant detuning of the laser frequency from the $F=3 \rightarrow F'=4$ transition determined the selected velocity, $v=5.0\times10^{2}$~m/s. The width of the selected velocity distribution was estimated at $\sim5-10$~m/s, which was primarily determined from the natural linewidth (6~MHz) of the $F=3 \rightarrow F'=4$ transition.

\section{Magnetic lattice}

\begin{figure}
\includegraphics{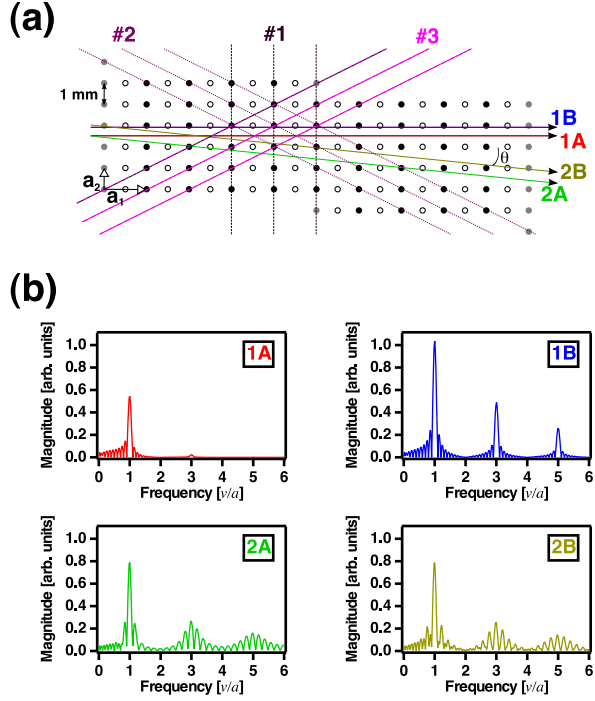}
\caption{(color online). (a) Configuration of wire current. The wires ran normal to the paper. Solid and open circles represent the positions of wires through which current flowed forward and backward, respectively. Grey solid circles at the edges of the array correspond to the wires for which current was halved compared to the other wire currents. The wire size (diameter: 30~$\mu$m) is magnified 10 times in the figure. Four examples of the beam trajectories are represented by the arrows 1A, 1B, 2A, and 2B. $\theta$ represents the angle between the channeling plane (the plane of the horizontal array of wires) and the atomic beam. The lattice axes $\mathbf{a}_1$ and $\mathbf{a}_2$ are shown. The arrays of dashed lines, labeled as \#1, \#2 and \#3, indicate the arrays of lattice planes causing the resonance. (b) The Fourier transform spectra (magnitude) of the transverse magnetic field (perpendicular to the atomic beam) experienced by atoms as a function of frequency in the unit of the atom velocity $v$ divided by the field period $a$. The Fourier analysis was performed assuming an array of infinitely long wires. The graphs are labeled using the names of the beam trajectories depicted in (a).}
\end{figure}

The magnetic lattice was produced by current flowing through approximately seven layers of planar arrays of parallel wires. The current configuration is shown in Fig.~2(a). The wire array was homemade. Gold-coated tungsten wires with a diameter of 30~$\mu$m spanned printed-circuit boards (PCBs) that were separated by 30~mm and formed the side walls of the wire array. Each wire was soldered at through-holes of both PCBs with some tension applied. The wires were connected in series on the PCBs, and side-by-side currents flowed parallel in opposite directions, generating a periodic magnetic field. The separation between adjacent wires was 1.0~mm, and the field period was 2.0~mm. Altogether, 129 wires were used to produce the magnetic lattice with 10 cycles of 2-mm periods and a $\sim5$~mm$\times$30~mm beam entrance, which was large enough to accept the atomic beam. Note that currents at the edges of the arrays were halved by connecting pairs of edge wires in parallel to compensate for the distortion of the periodic field produced by a finite array of current-carrying wires~\cite{Sid96, Kob09}. 

To suppress the dephasing of spin nutation for an ensemble of atoms in the beam, it is necessary for all atoms to experience the same amplitude from the periodic field. Otherwise, the nutation signal would decay quickly due to inhomogeneous spin nutation frequencies. This condition, however, was not fulfilled when the incident beam was parallel to the planar arrays of parallel current-carrying wires (referred to as ``channeling planes''~\cite{Gem74} hereafter), as shown in the two trajectories 1A and 1B in Fig.~2(a). The amplitudes of the periodic field are larger for atoms traveling closer to the wire planes. This fact is illustrated in Fig.~2(b), which shows the Fourier transform spectra of the periodic field experienced by atoms in the trajectories labeled in Fig.~2(a). The upper-left graph corresponds to the trajectory 1A at the center between two channeling planes, and the upper-right figure corresponds to the near-channeling-plane trajectory 1B. The main Fourier component at a frequency of $v/a$ was stronger in the latter than in the former. In addition, the Fourier components at third and fifth orders in $v/a$ were also prominent in the latter. The small sideband components originated from the transit-time effect, which also determined the linewidth.

The dependence of the field amplitude on the beam trajectory can be resolved by using ``oblique'' incidence. The graphs labeled 2A and 2B in Fig.~2(b) show the Fourier transform spectra for atoms incident upon the lattice with $\tan\theta=1/10$, where $\theta$ is the angle between the channeling plane and the atomic beam (the atoms traveled across the channeling planes twice before they exited the lattice). The main Fourier components at $v/a$ had the same amplitude in both graphs, regardless of the beam incident position. 
The relatively large sideband components were produced by other Fourier components corresponding to other periodicities of the lattice (see below).

The above discussion can be understood in terms of crystal physics~\cite{Kit05,Kon06}. A periodic field $\mathbf{B}(\mathbf{r})$ that is a function of position $\mathbf{r}$ can be expanded in a Fourier series 
\begin{eqnarray}
\mathbf{B}(\mathbf{r})=\sum_{\mathbf{G}} \mathbf{B_G}\exp(i\mathbf{G}\cdot\mathbf{r}), 
\label{Fourier}
\end{eqnarray}
where $\mathbf{G}$ is the reciprocal lattice vector, and $\mathbf{B_G}$ is the Fourier coefficient. We defined the crystal axes as $\mathbf{a}_1$ and $\mathbf{a}_2$, as indicated in Fig.~2(a); $|\mathbf{a}_1|=2|\mathbf{a}_2|=a$. $\mathbf{G}$ is expressed as $\mathbf{G}=u_1\mathbf{b}_1+u_2\mathbf{b}_2$,
where $\mathbf{b}_1=(2\pi/a^2)\mathbf{a}_1$ and $\mathbf{b}_2=(8\pi/a^2)\mathbf{a}_2$ are the axis vectors of the reciprocal lattice, and $u_1$ and $u_2$ are integers.
Assuming a straight classical beam trajectory of $\mathbf{r}=\mathbf{v}t+\mathbf{r}_0$, where $\mathbf{r}_0$ represents the beam position when it entered the lattice at time $t=0$, and substituting it in the above equation, one understands that the Fourier component specified by $\mathbf{G}$ formed the ``plane wave'' $\mathbf{B_G}\exp(i(\mathbf{G}\cdot\mathbf{v}t+\mathbf{G}\cdot\mathbf{r}_0))$. 
For example, the $(u_1, u_2)=(1, 0)$ component forms the plane wave with a frequency of $(v/a)\cos\theta \sim v/a$ (when $\theta \ll 1)$, corresponding to the main frequency component.
The incident position affected only the phase factor, $\exp(i\mathbf{G}\cdot\mathbf{r}_0)$, and not the amplitude $|\mathbf{B_G}|$.
The $(u_1, u_2)=(1, \pm1)$ components had frequencies of $(v/a)(\cos\theta \mp 2\sin\theta)$, respectively, corresponding to two sidebands near the main peak observed under the oblique incident condition. On the other hand, when the incident beam was parallel to the channeling planes ($\theta=0$), Fourier components with the same $u_1$ but different $u_2$ had the same frequency $u_1v/a$. In this degenerate case, many $|\mathbf{B_G}|$ values added with the phase factors $\exp(i\mathbf{G}\cdot\mathbf{r}_0)$, resulting in the dependence of the amplitude on the incident position at frequency $u_1v/a$.

It is interesting that the plane wave generated by one Fourier component was ``linearly polarized'': each Fourier coefficient of the magnetic field $\mathbf{B_G}$ was perpendicular to the wire and $\mathbf{G}$. This relation was derived when we Fourier-expanded the vector potential $\mathbf{A}(\mathbf{r})$, as in Eq.~(\ref{Fourier}). For an array of infinitely long straight currents, $\mathbf{A}(\mathbf{r})$ and $\mathbf{A_G}$ had only the component parallel to the current direction. Because $\mathbf{B}=\nabla\times\mathbf{A}$, $\mathbf{B_G}$ was parallel to $\mathbf{G}\times\mathbf{A_G}$ (that is, it was perpendicular to the wire and $\mathbf{G}$).

Because one reciprocal lattice vector specifies one array of lattice planes so that it is perpendicular to the lattice planes, and its magnitude divided by $2\pi$ equals the reciprocal of interplanar separation, each frequency component was generated by the corresponding array of lattice planes across which the atoms traveled. For example, the $(u_1, u_2)=(1,0)$ Fourier component was generated by the array of lattice planes labeled as \#1 in Fig.~2(a), whereas the $(u_1, u_2)=(1, 1)$ and $(1, -1)$ components were generated by the \#2 and \#3 planes, respectively. 

\section{Results and Discussion}

\begin{figure}
\includegraphics[width=6cm]{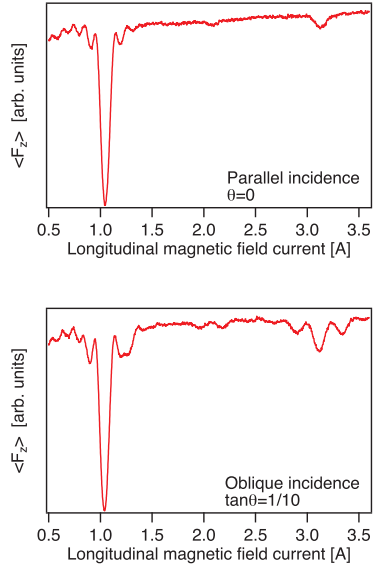}
\caption{(color online). Magnetic resonance spectra. The spin component $\langle F_z\rangle$, measured after the beam exited the lattice, is shown as a function of the coil current producing the longitudinal magnetic field. The upper graph represents the spectrum for the atomic incident beam that was parallel to the channeling planes; the lower graph represents the oblique incident beam with $\tan\theta=1/10$. The coil current produced the longitudinal magnetic field at $4.9\times10^{-2}$~mT/A. }
\end{figure}

Figure~3 shows magnetic resonance spectra measured as a function of the coil current producing the longitudinal magnetic field. The lattice current was fixed at 6.95~mA, which approximately fulfilled the $\pi$ pulse condition for the main resonance at a transition frequency of $v/a$, as shown below. The upper graph represents the spectrum for the atomic incident beam that was parallel to the channeling planes, whereas the lower graph represents the oblique incident beam with $\tan\theta=1/10$. 
The overall trend of the magnetic resonance spectra can be understood by following the discussion in the previous section. Both figures show strong resonance peaks at 1.05~A. At this current, the Zeeman splitting was estimated at $2.4\times10^2$~kHz, which is consistent with the main resonance frequency, $v/a$. The resonance peaks at the third order in $v/a$ were also clearly seen at 3.1~A. The peaks corresponding to the second order resonance were weak but observable. Although their observation was not expected based on the theoretical Fourier analysis shown in Fig.~2(b), it was likely a result of imperfections in the homemade wire array because the other array we made did not show second order resonance peaks. Many sideband peaks were also observed, as predicted in Fig.~2(b). 

\begin{figure}
\includegraphics[width=6cm]{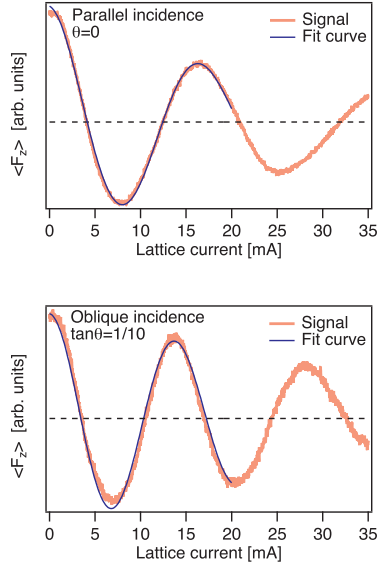}
\caption{(color online). Spin nutation signals. The spin component $\langle F_z\rangle$ is shown as a function of the lattice current producing the magnetic lattice. The longitudinal magnetic field current was fixed at the main resonance. The upper graph represents nutation for the atomic incident beam parallel to the channeling planes; the lower graph represents the oblique incident beam with $\tan\theta=1/10$. The experimental data were fit with exponentially decaying cosine functions. The dotted lines represent the baselines of the nutation determined from the fittings.}
\end{figure}

We fixed the longitudinal field current at the main resonance peak and measured spin nutation, which evolved as the atoms passed through the lattice. Figure~4 shows the spin component $\langle F_z\rangle$ as a function of the lattice current for the parallel incident beam (upper graph) and for the oblique incident beam with $\tan\theta=1/10$ (lower graph). For both beams, as the lattice current and the strength of the periodic field increased, the spin nutated further, and clear spin nutation signals were observed. The signals fit well with exponentially decaying cosine functions. The nutation rate of 2$\pi$~rad/14.0~mA was derived from the fit for the oblique incidence and was consistent with the nutation frequency estimated from the Fourier analysis: the transverse field amplitude 0.0110~mT at 14.0~mA generated a spin nutation of 25.7~kHz (under the rotating wave approximation), which was the reciprocal of the transit time. The nutation for the oblique incidence decayed 2.3 times slower (in units of nutation cycles) than that for the parallel incidence. This was likely due to the fact that the field amplitude was independent of the incident position of the oblique incidence, as expected. We found that the other homemade wire array gave similar nutation signals with slightly different decay rates. Therefore, we considered that the observed dephasing was, in part, caused by the imperfection of the lattice.

\section{Conclusions}

In this paper, we demonstrated that spin nutation was induced by atomic motion through a magnetic lattice. This was the first demonstration of coherent population transfer in RCE or MIR experiments, which used a periodic structure to induce resonance transitions for atoms in a beam. The magnetic lattice was generated by current flowing through a multiwire-chamber-like array of parallel wires, which allowed a variety of incident beam angles. Motion-induced resonance occurred at frequencies predicted by Fourier analysis of the lattice field. The atomic spin nutated as the atoms passed through the lattice under the resonance condition, and a resultant sinusoidal variation in the spin component $\langle F_z\rangle$ was clearly observed as a function of the strength of the lattice field. Slow decay of coherent oscillation in nutation signals was obtained for oblique incident atoms, which experienced the same field amplitude regardless of their incident positions. Our demonstration of spin nutation showed that coherent manipulation of atoms, one of the most useful capabilities of atom--photon interactions, can also be realized by RCE or MIR.

\begin{acknowledgments}
This work was supported by KAKENHI (No. 20684017) and the ``Improvement of Research Environment for Young Researchers'' program of The Ministry of Education,
 Culture, Sports, Science, and Technology, Japan.
\end{acknowledgments}

%\bibliography{MIR_RCE}

%merlin.mbs 2010-03-15 4.21a (PWD, AO, DPC)

\end{document}